\documentclass[manuscript]{acmart}
\AtBeginDocument{%
  }

\setcopyright{acmlicensed}
\copyrightyear{2026}
\acmYear{2026}
\acmDOI{XXXXXXX.XXXXXXX}
\acmConference[PEARC '26]{Make sure to enter the correct
  conference title from your rights confirmation email}{July 26--30,
  2026}{Minneapolis, MN}
\acmISBN{978-1-4503-XXXX-X/2026/07}

\usepackage{multirow}
\usepackage{graphicx}
\usepackage{caption}
\usepackage{subcaption}

\usepackage{xcolor} 
\usepackage{soul}   

\begin{document}

\title{A Precision Emulation Approach to the GPU Acceleration of Ab Initio Electronic Structure Calculations}

\author{Hang Liu}
\email{hliu@tacc.utexas.edu}
\orcid{0000-0002-3486-7863}
\author{Junjie Li}
\email{nicejunjie@gmail.com}
\orcid{0000-0002-1051-5927}
\author{Yinzhi Wang}
\email{yinzhi.wang.cug@gmail.com}
\orcid{0000-0001-8505-0223}
\affiliation{%
  \institution{Texas Advanced Computing Center, The University of Texas at Austin}
  \city{Austin}
  \state{Texas}
  \country{USA}
}
\author{Niraj K. Nepal}
\email{nnepal@psc.edu}
\orcid{0000-0002-7281-3268}
\author{Yang Wang}
\email{ywg@psc.edu}
\orcid{0000-0002-9837-5796}
\affiliation{%
  \institution{Pittsburgh Computing Center, Carnegie Mellon University}
  \city{Pittsburgh}
  \state{PA}
  \country{USA}
}
\begin{abstract}
This study explores the use of INT8-based emulation for accelerating traditional FP64-based HPC workloads on modern GPU architectures. Through \texttt{SCILIB-Accel} automatic BLAS offload tool for cache-coherent Unified Memory Architecture, we emulate FP64 matrix multiplications in the LSMS CPU application in the \texttt{MuST} suite without code changes. We find that accuracy depends on both arithmetic precision and the properties of the operator, which can be dealt with through tunable precision emulation. Unlike traditional mixed-precision approaches, this method preserves original algorithms while optimizing hardware utilization. We showcase the potential of improving accuracy and performance at the same time. This work highlights the potential of AI-driven hardware to transform HPC, advocating for adaptive precision strategies in future scientific computing.
\end{abstract}

\ccsdesc[500]{General and reference~Performance}
\ccsdesc[500]{Mathematics of computing~Mathematical software performance}
\ccsdesc[500]{Software and its engineering~Software performance}

\keywords{GEMM, BLAS, Automatic Offload, FP64 Emulation, Mixed-Precision, Ozaki Scheme, Ab Initio, Electronic Structure Calculation}

\received{30 March 2026}

\maketitle

\section{Introduction}
Rapid advancement in artificial intelligence (AI) has triggered the development of specialized hardware, such as NVIDIA GPUs with Tensor Cores, AMD GPUs with Matrix Cores and various TPUs, to accelerate machine learning computations with high-throughput and low-precision arithmetic units\cite{abdelfattah_survey_2021}. 
These processors take advantage of low/mixed-precision floating-point formats (e.g., FP16, BF16) and low-bitwidth integer units (e.g., INT8, INT4) to increase performance, reduce power consumption, and minimize memory footprint.
While these architecture innovations have had a dramatic impact on AI training and inference performance, their potential to accelerate traditional HPC workloads that were historically based on CPUs and double-precision floating point (FP64) computations for scientific simulation, numerical modeling, and other compute-intensive tasks, remains vast yet not fully understood, requiring further research and development to unleash.

FP64 has been the norm in traditional HPC, ensuring the numerical precision required for simulations like climate modeling, quantum chemistry simulation, and computational fluid dynamics. 
However, current hardware market has been heavily influenced by the AI mania which strongly favors low-precision data formats, resulting reduced or even nearly diminished FP64 capability in some of the recent GPUs such as NVIDIA's Blackwell series and Rubin \cite{nvidia-hgx}. 
Therefore academic HPC centers serving the broad research community have to balance the selection of AI-centric hardware and conventional HPC capable hardware.  
This scenario provides an opportunity to rethink computational strategies in HPC and harness the extreme power of low-precision data types. By offloading suitable workloads on GPUs and leveraging reduced-bit-width types like INT8, one can emulate FP64 computations at acceptable precision, promoting convergence of HPC and AI hardware and maximize the efficiency of hardware utilization. 

This work investigates the integration of INT8-based FP64 GEMM emulation and automatic BLAS offloading \cite{li_automatic_2024} into a conventional density functional theory (DFT) based {\it ab initio} electronic structure calculation package \texttt{MuST}~\cite{must-url}. 
By adapting the Ozaki framework, we leverage high-throughput INT8 arithmetic units on modern GPUs to achieve tunable precision across various levels of numerical fidelity. Our results demonstrate that the underlying algorithms of \texttt{MuST} are inherently compatible with GEMM emulation, providing systematic control over physical accuracy while delivering significant performance gains. This study showcases the potential for maintaining high numerical accuracy without compromising throughput via automated offloading and tunable precision emulation on modern GPU architectures.

\section{Related Work}
\subsection{Previous Work on Automatic BLAS Offload}
The main limitation of GPU acceleration in traditional HPC codes is the difficulty of retrofitting legacy CPU applications. Manual porting is labor-intensive and impractical for most scientific communities. \texttt{MuST} studied in this work is one such case as a CPU-native code, achieving a high-performance GPU port is highly challenging.

Several efforts have aimed to automatically offload BLAS calls from CPUs to GPUs. Cray \texttt{LIBSCI\_ACC} \cite{cray-libsci-acc, cray-libsci-acc-slide} pioneered this approach, IBM’s \texttt{ESSL} \cite{ibm-essl-offload} and NVIDIA’s \texttt{NVBLAS} \cite{nvidia-nvblas} works similarly. 
These tools replace CPU BLAS calls with GPU BLAS calls, and automate data movement and offload decisions. However, they target conventional GPU architecture, and data is transferred between CPU and GPU for each offloaded BLAS call, leading to unacceptable high overhead and significant performance loss in practical HPC applications \cite{li_automatic_2024, li_automatic_2025}.

A significant advancement is presented by Li et al. \cite{scilib-accel,li_automatic_2024,li_automatic_2025}, who developed a high-performance automatic BLAS offloading tool, \texttt{SCILIB-Accel}. It introduces a novel data movement strategy that leverages cache coherency in recent CPU–GPU systems, such as NVIDIA Grace-Hopper, and AMD GPUs with cache-coherent Infinity Fabric.
The new data movement strategy is designed to maximize reuse in common matrix-math-based scientific workloads. Evaluated on real HPC applications, the tool delivers substantial performance gains without code changes, overcoming the limitations of prior approaches and setting a new standard for BLAS offloading in HPC.


\vspace{-0.1cm}
\subsection{GEMM Emulation on Integer Matrix Multiplication Unit}
The study of floating point emulation isn't new. 
The most widely known emulation is emulating quad-precision using double-precision \cite{hida_algorithms_2001} for those scientific applications that mandates precision beyond FP64. 
In recent years, a series of papers and implementations on the Ozaki schemes
were published exploring methods for emulating high-precision matrix multiplication (e.g., FP64/double precision) using lower-precision hardware (e.g., INT8/FP8 Tensor Cores) while maintaining the accuracy of the higher precision. It is a generic scheme for matrix multiplications by decomposing high-precision matrices into multiple low-precision components, performing multiple matrix multiplications, and reconstructing the result.
The original Ozaki-I splits high-precision input matrices into slices as lower-precision components based on their significant bits and exponent alignment, then performs low-precision matrix multiplications on these slices and accumulates them in higher precision\cite{ootomo_dgemm_2024, doi:10.1177/10943420241313064}. Its open source implementation \texttt{ozIMMU} and improvements \texttt{ozIMMU-H} are available at \cite{ozIMMU-url} and \cite{ozIMMU-H-url}. The \texttt{cuda13} in \texttt{NVHPC/26.1} SDK includes the Nvidia implementation based on Ozaki-I \cite{cuda13-emulation-url}.
The later developed Ozaki-II leverages the Chinese Remainder Theorem (CRT) for matrix multiplication emulation~\cite{uchino2025emulationcomplexmatrixmultiplication}. It converts floating-point matrices into integers, performs multiple matrix multiplications using smaller, pairwise coprime moduli and uses the CRT to reconstruct the final result. The Ozaki-II is generally proven higher performance, more power efficient and better accuracy control than those of Ozaki-I on modern hardware, e.g., NVIDIA GB200, that features much faster INT8/FP8 Tensor Cores. Its open source implementation \texttt{GEMMul8} is available at \cite{gemmul8-url}.

\section{Experiment}
\subsection{Implementation}
In order to run a CPU code on GPU's INT8 tensor cores, our experiments use two open-source tools: \texttt{SCILIB-Accel}~\cite{scilib-accel,li_automatic_2024,li_automatic_2025} and \texttt{GEMMul8}~\cite{gemmul8-url}. The \texttt{GEMMul8} hook mode redirects \texttt{cuBLAS} GEMM function calls to its \texttt{GEMMul8} implementation using the dynamic linker, enabling the INT8-based GEMM emulation. The \texttt{SCILIB-Accel}, utilizing the \texttt{PEAK} profiler framework~\cite{wang_peak_2023}, employs DBI to transparently intercept BLAS calls and automatically offload them to GPU through calling \texttt{cuBLAS}. The transparent DBI implementation by design allows both tools to operate collaboratively with a single \texttt{LD\_PRELOAD} to achieve automatic offloading. This approach enables tunable precision emulations on any GEMM-heavy CPU programs without code change or recompiling. 
The emulation precision of GEMM is primarily controlled by the moduli numbers set by environment variables~\cite{gemmul8-url}. 
Our experiments also use the Ozaki-I implementation in \texttt{cuda13}.  
It provides environment variables to control the emulation strategy and precision without code change. Using it just needs to build the \texttt{SCILIB-Accel} by \texttt{cuda13}~\cite{cuda13-emulation-url}. 

\subsection{Experiments with MuST}

\texttt{MuST} (Multiple Scattering Theory) \cite{must-url} is an {\it ab initio} software package for electronic structure calculations that solves the Kohn–Sham equation using the Green function method. The Locally Self-consistent Multiple Scattering (LSMS) method~\cite{must_1995} implemented in \texttt{MuST} is specifically targeted for large-scale systems, featuring linear scaling with respect to the number of atoms ($O(N)$). The LSMS code is a two-time recipient of the Gordon Bell Prize, awarded in 1998 and 2009 for its groundbreaking performance in high-performance computing \cite{10.5555/509058.509129,gordonbell09}. The LSMS method relies heavily on BLAS operations, primarily the \texttt{ZGEMM} and \texttt{ZTRSM}, which often account for over 80\% of CPU runtime.

Since the primary computational bottleneck of the LSMS method in \texttt{MuST} is the LU-based inversion of the multiple scattering matrix evaluated across a complex energy contour, its high \texttt{ZGEMM} intensity makes it an ideal testbed for assessing emulation accuracy. Within the multiple scattering formalism, the Green function $G(\mathbf{r},\mathbf{r};z)$ of the Kohn-Sham equation, $\left[-\nabla^2+V_{\text{eff}}(\mathbf{r})\right]\psi(\mathbf{r};z)=z\,\psi(\mathbf{r};z)$, is determined through this matrix inversion. The valence electron density is then obtained by performing an integration of the Green function along a complex contour $C$ in the upper half plane which originates below the valence band and terminates at the Fermi energy: $\rho({\bf r})=-\frac{1}{\pi}{\rm Im}\int_C G({\bf r},{\bf r};z)\,dz$. In practice, this contour integration is typically performed over a semi-circular path using a $\sim$30-point Gaussian quadrature. 
Since the accuracy of the electron density and all other observables is fundamentally tied to the accuracy of the Green function, we investigated the percent error in the energy-dependent, atomic volume ($\Omega$) integrated Green function 
$G(z)=\int_\Omega{G({\bf r},{\bf r};z)\,d{\bf r}}$ across various emulation precision modes in \texttt{cuda13} and \texttt{GEMMul8}, ranging from the \texttt{cuda13} eager strategy with mantissa bits 31, 39, 47, 55, 63, to the \texttt{GEMMul8} moduli numbers 10, 12, 14, 16, 18, and used the native FP64 GEMM as our ground-truth baseline. This serves as the most sensitive and critical metric to assess the fidelity of different emulation modes. We then monitored three primary physical observables, all-electron total energy per atom $E_{\rm tot}$, local magnetic moment $\mu$, and local net electronic charge $\delta Q$. These quantities are functionals of the electron density of both spin up and down states. 
We selected the FeNi\textsubscript{3} benchmark from the \texttt{MuST} distribution. This case performs a noncollinear magnetism study of FeNi\textsubscript{3} alloy (in L1\textsubscript{2} crystal structure) using the LSMS method.  
The CPU version of \texttt{MuST}, \texttt{SCILIB-Accel} and \texttt{GEMMul8} are built by \texttt{nvhpc/26.1} with the \texttt{nvhpc-hpcx-cuda13/26.1} module loaded. The benchmarks were performed on a Nvidia GB200 partition on a NVL4 node.


\section{Discussion}
\begin{figure}[h!]
  \centering
  \begin{subfigure}[b]{0.42\textwidth}
    \includegraphics[width=\linewidth]{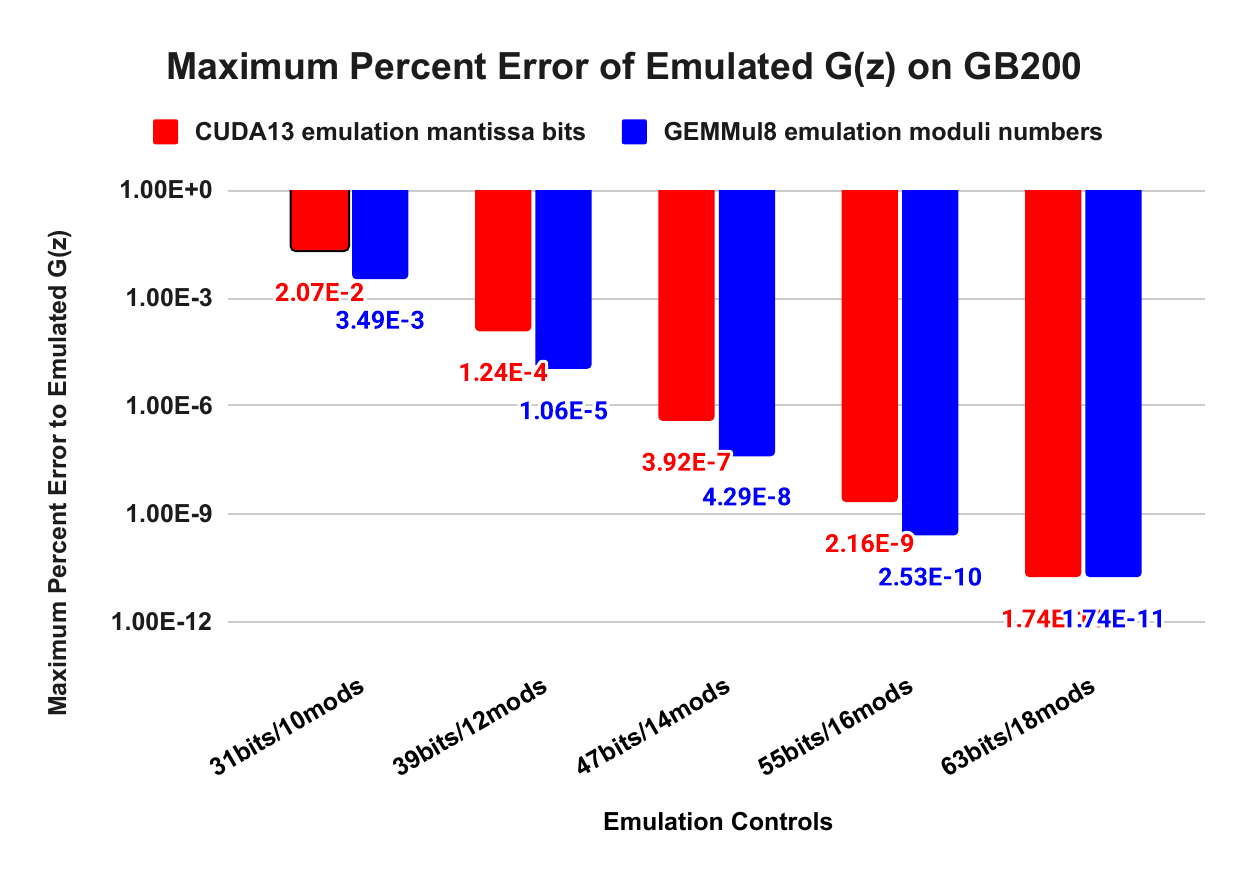}
    \vspace{-0.5cm}
    \caption{Maximum percent error of emulated $G(z)$.}
    \label{fig:GF_error}
  \end{subfigure}
  \hspace{1cm}
  \begin{subfigure}[b]{0.42\textwidth}
    \includegraphics[width=\linewidth]{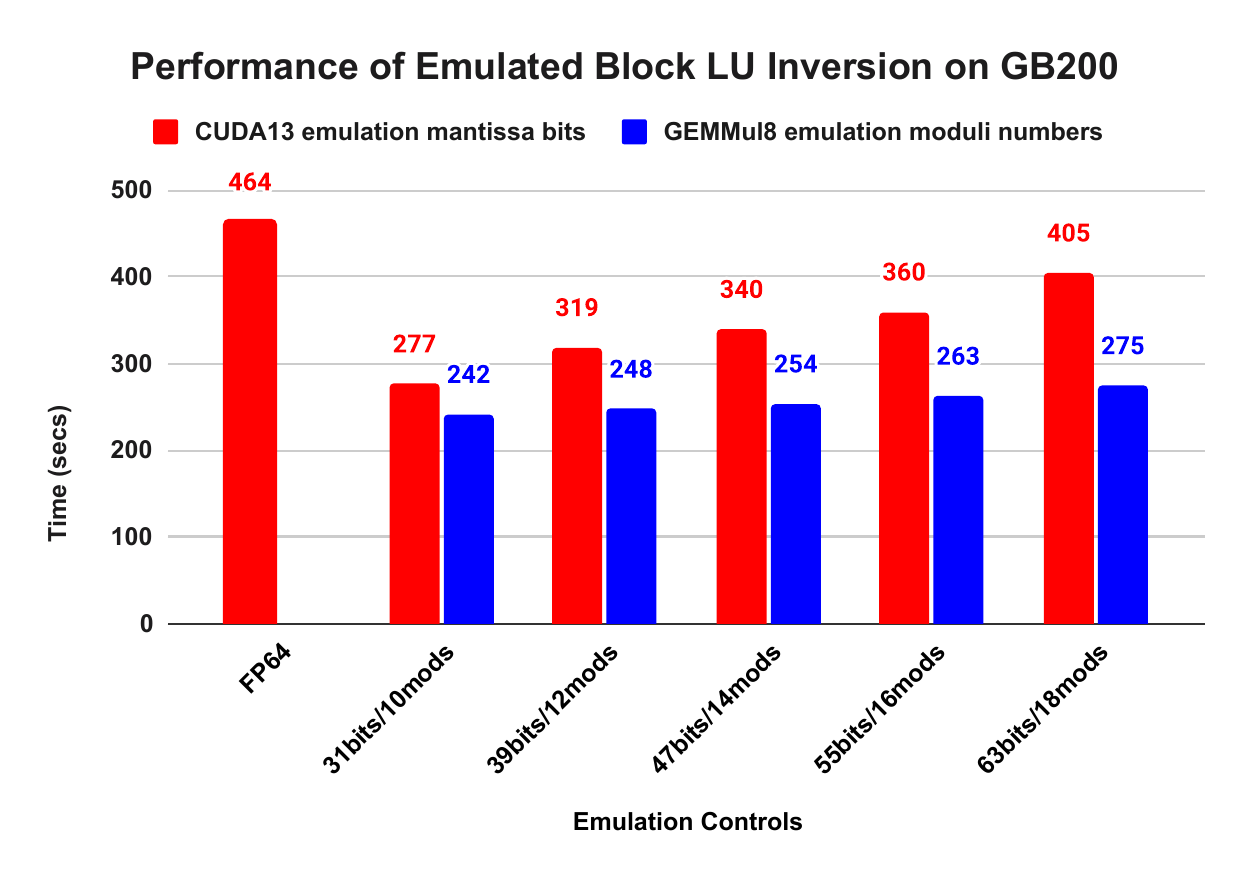}
    \vspace{-0.5cm}
    \caption{Cumulative timing of the Green function calculations.}
    \label{fig:GB200_perf}
  \end{subfigure}
  \vspace{-2.0mm}
  \caption{
    Impact of mixed-precision emulation on the LSMS method using cuda13 and GEMMul8. The $G(z)$ counts all energy points and atomic sites, requiring invertion of a $33,750 \times 33,750$ double complex matrix, representing a significant computational challenge for standard FP64 solvers.}
  \label{fig:max_perf}
\end{figure}

To establish the baseline, we ran the MuST LSMS on CPU and CPU-GPU offloading by \texttt{SCILIB-Accel} with native FP64 mode.
Figure.\ref{fig:GF_error} illustrates the maximum percent error of $G(z)$ evaluated for all Gaussian quadrature energy points under different emulation modes against the native FP64 mode. The data label, $i$ bits/$j$ mods, marks different emulation modes: the $i$ is the mantissa bits count in \texttt{cuda13} eager strategy, the $j$ is the moduli numbers in GEMMul8. The lower precision 31bits/10mods mode has a maximum percent error up to $10^{-2}$. Then the 39bits/12mods and 47bits/14mods modes demonstrate exponential improvements. The 55bits/16mods mode achieves accuracy up to $10^{-10}$, which is a sufficient level of variance typical of FP64 code across different compilers or HPC systems. Although the 63bits/18mods mode may offer even higher accuracy, its impact is
limited by the rest FP64 codes in the \texttt{MuST} package. The corresponding cumulative execution time of each emulation mode for the scattering matrix inversion
of 2 SCF iterations is listed in Figure.\ref{fig:GB200_perf}. It shows the accelerated inversion of FP64 emulation on INT8. Especially, the averaged 1.7X speedup by the \texttt{GEMMul8} high precision modes underscores the effective trade-off between numerical accuracy and the high-throughput capabilities of INT8 Tensor Cores.

Table. \ref{tab:physical_observable_error} summarizes the impact of emulation controls to the accuracy of physical observables. 
Importantly, by running the SCF calculation with the same starting potential, all higher-precision emulation modes achieved self-consistency for the effective potential and electron density within $10^{-6}$ and matched the numerical accuracy of the FP64 baseline, whereas the 31-bit mode failed to converge. 
It is noteworthy that the total energy from 31bits mode also demonstrates high fidelity to the FP64 as well, even the $G(z)$ percent error is up to $10^{-2}$ at that mode. This stability is expected in the LSMS method. 
\begin{table}[h!]
\vspace{-0.25em}
\centering
\caption{Impact of Emulation Controls on Accuracy of Physical Observables. The total energy per atom $E_{\rm tot}$ is in the units of Rydberg, the local magnetic moment $\mu$ is in the unit of Bohr magneton. The PBE exchange-correlation functional is employed in the calculation.}
\vspace{-2mm}
\resizebox{\textwidth}{!}{
\begin{tabular}{c|c|c|c|c|c|c|c|c|c}
\hline
 mode  & $E_{\rm tot}$  & $\mu({\rm Fe})$ & $\mu({\rm Ni}_1)$ & $\mu({\rm Ni}_2)$ & $\mu({\rm Ni}_3)$ & $\delta Q({\rm Fe})$ & $\delta Q({\rm Ni}_1)$ & $\delta Q({\rm Ni}_2)$ & $\delta Q({\rm Ni}_3)$ \\
\hline
\texttt{GPU FP64$^{*}$} & -2894.19443 & 3.00418 & 0.58427 & 0.58427 & 0.58427 & -0.29313 & 0.09771 & 0.09771 & 0.09771 \\
\texttt{cuda13 31bits} & -2894.19443 & 3.00417 & 0.58426 & 0.58429 & 0.58425 & -0.29313 & 0.09771 & 0.09769 & 0.09772 \\
\texttt{cuda13 55bits} & -2894.19443 & 3.00418 & 0.58427 & 0.58427 & 0.58427 & -0.29313 & 0.09771 & 0.09771 & 0.09771 \\
\texttt{GEMMul8 10 moduli} & -2894.19443 & 3.00417 & 0.58427 & 0.58427 & 0.58427 & -0.29313 & 0.09771 & 0.09771 & 0.09771 \\
\texttt{GEMMul8 16 moduli} & -2894.19443 & 3.00418 & 0.58427 & 0.58427 & 0.58427 & -0.29313 & 0.09771 & 0.09771 & 0.09771 \\
\hline
\end{tabular}
}
\begin{minipage}{\textwidth}
\raggedright\footnotesize
\vspace{4pt}
* GPU FP64 mode is CPU binary combined with SCILIB-Accel automatic BLAS GPU offload.
\end{minipage}
\label{tab:physical_observable_error}
\vspace{-14pt}
\end{table}
First, the numerical stability of the LSMS method is supported by the spectral characteristics of $G(z)$. As shown in our previous pilot study \cite{pilot_study_2025}, significant percent errors stemming from emulation are primarily confined to a narrow region near the Fermi energy, where the Green function exhibits sharper spectral features. However, because each Gaussian quadrature point along the complex contour $C$ carries a relatively small integration weight, these localized emulation errors are effectively suppressed and averaged out during the contour integration. Second, consistent with the variational principle in DFT-based ab initio methods, the total energy is stationary with respect to the electron density. Consequently, a first-order error in the density resulting from lower-precision emulation modes leads only to a second-order error in the total energy. 
This observed robustness against emulation quantization noise highlights the suitability of the LSMS framework to the GPU acceleration particularly when leveraging emulation-based tensor core operations.

\section{Conclusion}

In this study, we presented a preliminary investigation into emulating floating-point matrix multiplication operations using INT8 data types. This approach, combined with our previously developed automatic BLAS GPU offloading tool, is applied to the well-known FP64-based quantum physics CPU code, LSMS in the \texttt{MuST} suite. 
Using the implementation of Ozaki schemes,  
we explored various emulation modes to identify the optimal balance between efficiency and the necessary accuracy for scientific simulations. 
Our results demonstrate that these emulation schemes are both effective and accurate for FP64-based scientific applications, enabling accelerated simulations to be carried out without sacrificing accuracy.

Our emulation approach offers a broader and generic solution for scientific simulations, distinguishing itself from the typical mixed-precision studies, where specific solver algorithms are modified to accommodate lower precision in intermediate computations. 
Unlike those approaches, our method preserves the integrity of the original algorithm and code without modification while optimizing the utilization of hardware resources. 
We advocate for a re-evaluation of precision requirements in scientific applications and determine the precision that are truly necessary. 
We encourage the adoption of low-bitwidth types to enhance hardware utilization, particularly in the context of the growing influence of AI on new hardware. 
Additionally, we urge collaboration between hardware developers and computational scientists to design optimal data types that can better meet the demands of future scientific computing along with AI.

\begin{acks}
This work is supported by the National Science Foundation under Award Nos. OAC-2402542 and OAC-2139536, as well as the Leadership Class Computing Facility (LCCF) Characteristic Science Application project. The authors thank the NVIDIA Corporation for providing access to an NVL4 node, which is architecturally similar to the nodes of the LCCF Horizon system. Additionally, the authors acknowledge the technical assistance of AI tools (Grok, ChatGPT, Claude, and Gemini) in the drafting and linguistic refinement of this manuscript.
\end{acks}

\bibliographystyle{ACM-Reference-Format}
\bibliography{reference}

\end{document}